\documentclass[acmsmall,screen,nonacm]{acmart}

\AtBeginDocument{%
  }

\setcopyright{none}
\renewcommand\footnotetextcopyrightpermission[1]{}

\usepackage{booktabs}
\usepackage{xcolor}
\usepackage{subcaption}
\usepackage{multirow}
\usepackage{xspace}
\usepackage{xfrac}
\usepackage{adjustbox}
\usepackage{wrapfig}
\makeatletter
\@ifclassloaded{acmart}{
}{
    \usepackage{amssymb}
}
\makeatother

\usepackage{algorithm}
\usepackage{algorithmic}
\usepackage{makecell}

\usepackage{newfloat}
\usepackage{listings}

\floatstyle{ruled}
\newfloat{listing}{tb}{lst}{}
\floatname{listing}{Listing}

\usepackage[textsize=tiny]{todonotes}
\setuptodonotes{inline}

\providecommand\mutt[0]{$\mathcal{MUT}$}

\providecommand\dolater[1]{}
\providecommand\hardcode[1]{#1}

\newcommand{\ProbeDropoutRate}{0.2\xspace}
\newcommand{\ProbeLearningRate}{0.001\xspace}

\newcommand{\ProbeNumEpochs}{30\xspace}

\newcommand{\GptNoSolvePct}{46\%\xspace}

\newcommand{\QwenNbSolvePct}{34\%\xspace}

\newcommand{\ComboPatchPct}{61\%\xspace}
\newcommand{\ComboModelPatchPct}{82\%\xspace}

\newcommand{\GptNoLineVerbalizedLinelevelUnscaledBss}{-0.18\xspace}
\newcommand{\GptNoLineVerbalizedLinelevelUnscaledEce}{0.26\xspace}

\newcommand{\QwenNbMultisamplingLinelevelScaledBss}{0.13\xspace}
\newcommand{\QwenNbMultisamplingLinelevelScaledEce}{0.04\xspace}

\newcommand{\QwenNbLineVerbalizedLinelevelUnscaledBss}{-0.72\xspace}
\newcommand{\QwenNbLineVerbalizedLinelevelUnscaledEce}{0.43\xspace}

\newcommand{\QwenNbProbeNbLinelevelScaledBss}{0.33\xspace}
\newcommand{\QwenNbProbeNbLinelevelScaledEce}{0.02\xspace}

\title{Localized Calibrated Uncertainty in Code Language Models}

\author{David Gros}
\email{dgros@ucdavis.edu}
\affiliation{%
  \institution{University of California, Davis}
  \city{Davis}
  \state{California}
  \country{USA}}

\author{Prem Devanbu}
\email{ptdevanbu@ucdavis.edu}
\affiliation{%
  \institution{University of California, Davis}
  \city{Davis}
  \state{California}
  \country{USA}}

\begin{abstract}
 Large Language models (LLMs) can generate complicated source code from natural language prompts. However, LLMs can generate output that deviates from what the user wants, requiring supervision and editing. To support this process, we offer techniques to localize where generations might be misaligned from user intent. 
  We first create a dataset of ``Minimal Intent Aligning Patches'' of repaired LLM generated programs. Each program uses test cases to verify correctness. 
  After creating a dataset of programs, we measure how well various techniques can assign a well-calibrated probability to indicate which parts of code will be edited in a minimal patch (i.e., give a probability that corresponds with empirical odds it is edited). 
  We compare white-box probing (where we propose a technique for efficient arbitrary-span querying), against black-box reflective and self-consistency based approaches. 
  We find probes with a small supervisor model can achieve low calibration error and Brier Skill Score of approx 0.2 estimating edited lines on code generated by models many orders of magnitude larger.
 We discuss the generalizability of the techniques, and the connections to AI oversight and control, finding a probe trained only on code shows some signs of generalizing to natural language errors if new probability scaling is allowed. 
\end{abstract}

\begin{document}

\maketitle

\section{Introduction}

Large language models can produce stunning amounts of complex text. 
Consider briefly a comparison to a skilled human typist, who, typing at full speed, might type 500 characters per minute, which might equate to roughly 10-20 lines of Python code per minute. 
Meanwhile, GPT-like models can produce thousands of lines of (mostly) syntactically
correct code per minute. 
While generated code may be buggy or wrong~\cite{jesseLargeLanguageModels2023b}, it could potentially save
time and effort. 
For example, in 2022, Google researchers reported that 3\% of company code involved AI generation \cite{tabachnyk2022ml}, but by 2025, this number has risen to 30\% \cite{alphabet2025q1}.

As LLM-generated code gets wider use and more complex, developers would benefit from a reliable estimate of how likely each part of a generated output sequence are to be correct (\emph{i.e.,} matching the user's intent). This could aid better deployment of quality control resources and might help reduce known safety risks of even capable models, via better oversight. 
A classical way to frame the estimation of uncertainty in predictive modeling is calibrated probabilities. 
This corresponds to predicting which parts of the output are, say, 40\% or 90\% confident, and having those probabilities correspond to likelihood of actual, empirical correctness. Calibrated probabilities would aid rational decision making under uncertainty, with better outcomes in expectation. Given the cost and challenges of reviewing code, having localized and well-calibrated probabilities (which can reliably indicate which parts of the generated code require the most attention) could be valuable. 


\dolater{When we examine GPT-4o generations for XXXX problems, the median incorrect solution was X lines long, }

In this paper we present several experiments studying this problem of producing calibrated, localized uncertainty, in the context of code generation. We study the following research questions:

\begin{enumerate}
    \item How well do the latent states of models represent calibrated localized uncertainty?
    \item Can the stochastic nature of models be used to produce calibrated localized uncertainty estimates via consistency between multiple samples from the model?
    \item How well can models reflect on their own outputs for localized uncertainty estimation?
    \item How well do the techniques generalize to new domains outside of software engineering?
\end{enumerate}

Each set of experiments shares a common evaluation methodology (described in Section~\ref{sec:locdataset}). 
We perform evaluations on data from both self-contained function generation data \cite{liuYourCodeGenerated2023a, jain2024livecodebench}, and generations of functions in highly stared GitHub projects \cite{repocod}.

\section{Motivation and Approach}

The primary goal of this paper is to help programmers produce better
quality code at lower cost by making better
use of LLM-generated code. 
The approach is to guide programmers
to focus their review and edit efforts on just those parts of generated code
which are most
likely to require changes. We accomplish this by 
exploring reliable \emph{localized} uncertainty values, at the line and
token levels. 
\begin{figure}[htbp]
    \centering
    \includegraphics[width=0.95\textwidth]{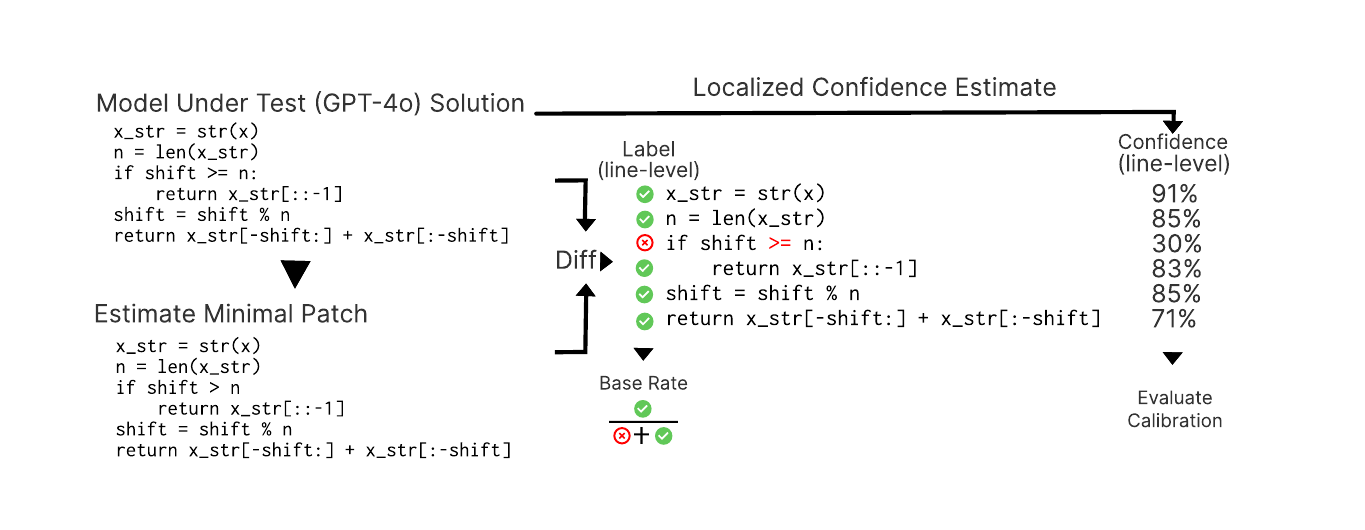}
    \caption{Illustrating the basic flow. GPT-4o generates a flawed solution with one line incorrect. 
    Our pipeline estimates the minimal patch. Then we take the diff which gives us labels at the token, line, and problem level.
    We illustrate sample confidences. From these we evaluate the quality of confidences via the Brier Score (MSE),
    which taking into account the underlying dataset base rate, becomes the Brier Skill Score (BSS), or ECE.
    }
    \label{fig:datasetfig}
\end{figure}

Figure~\ref{fig:datasetfig} illustrates our general approach. On the top-left is
the original generated code from GPT-4o\footnote{Code is part of HumanEval+ dataset, generated in response to the prompt ``Circular shift the digits of the integer x, shift the digits right by shift
and return the result as a string.
If shift > number of digits, return digits reversed." with some extra examples}. The generated code is not correct. Luckily, in this case, just one token ``{\small\tt{>=}}'', on 
the 3rd line requires repair. However, in current practice, all
the developer gets is the code.
While in this example we have  test cases ``behind the scenes'' to know it faulty, in many uses of LLM generated code, one only has the natural language prompt to go off of.
Thus it can be challenging 
to detect, diagnose, and localize the failure where the code does match intent. 

The right of the figure shows the same code,
but with confidence annotations, indicating likelihood of correctness.
Prior human-subject study by \citet{vasconcelos2023Exp} suggests that developers do indeed benefit from correct, localized indication of 
code likely in need of repair. 
Considering the public claims of over significant amount of code (at Google) being generated 
from LLMs \cite{alphabet2025q1}, 
focusing
of review/repair efforts, of potentially imperfect generated code, via reliable confidence measures, could prove valuable in the long run. 

We now briefly  outline  the rest of the paper. First, we begin in~\autoref{sec:related} with a survey of related work, and describing our main measures of ``calibration'', \emph{viz.,} the reliability of confidence
measures. 
Then, we describe our research method, which we amplify upon in subsequent sections. The first challenge we address (\autoref{sec:gathering}) is the task of building a suitable dataset  of incorrect 
generated code, and minimal applicable associated repair. Following this we describe different approaches
to computing fine-grained calibrated confidences associated with lines and tokens. We begin with our baseline raw probabilities from the model (\autoref{sec:baseline}) and then present a probing approach (\autoref{sec:probing}). Then we describe a multi-sampling approach (\autoref{sec:multisampling}), and our reflective approach (\autoref{sec:reflective}). Finally, 
we explore generalizability to natural language output, discuss the implications
of our work.

\section{Related work \& Background}
\label{sec:related}
We discuss some relevant concepts and prior work. 

\vspace{0.04in}

\noindent{\emph {Calibration \& its measurement:}}
When a prediction is probabilistically uncertain (\emph{e.g.,} ``Chance of rain tomorrow is 30\%"), it
\emph{could} still be used for rational decision-making, to improve the expected value of decisions. 
However, to support such decision-making, the confidence (\emph{viz.,} probability) value must be \emph{well-calibrated}: \emph{the probability number must reflect empirical frequency}, so that decisions made on the basis of the confidence value are likely to pay out in practice. Thus, on average, over all the times when the chance of rain is predicted to be $X\%$, it should rain almost exactly $X\%$ of those times. The higher degree
to which the predicted confidence matches the empirical frequency, the better ``calibrated" the predicted
confidence. Calibration for predictions from neural networks has long been a topic of interest~\cite{guoCalibrationModernNeural2017b,mindererRevisitingCalibrationModern2021a}.

To measure calibration we use two common measures of calibration, \emph{Brier score, and ECE}. 
Glenn Brier proposed the widely-used Brier score in 1950~\cite{brierVerificationForecastsExpressed1950a},
which measures the mean-squared error between confidence and the empirical outcome (\emph{e.g.,}, the value 1 if outcome is Rain, in the above example, and 0 otherwise) over a sample of predictions; it provides a
sample-statistic estimate of how far the predicted confidences deviates from the actual outcome. The calculation of the model's Brier score (smaller is better) on a sample  is seen in~\autoref{def:brier}: $p_i$ refers to the confidence in the $i^{th}$ prediction in a sample $T$ of size $N$, where $o_i$ is the outcome.

\begin{equation}
B_{model} = \frac{1}{N} \sum_{i=1}^{N} (\hat{p_i} - \hat{o_i})^2
\label{def:brier}
\end{equation}
\begin{equation}
B_{ref} = p *(1-p)
\label{def:briarref}
\end{equation}
\begin{equation}
BSS = \frac{(B_{ref} - B_{model})}{B_{ref}} 
\label{define:SS}
\end{equation}

We note that it's easier
to get a low (good) Brier score when the base-rate (\emph{e.g.,} the overall empirical frequency of rain) is really high or really low: in a rain forest, where it rains almost every day, a perpetual, daily 95\% prediction of rain probability will be mostly right, and get a low Brier.  Thus, for base rate $p$, a naive model could always predict probability $=p$; for such a model, one can calculate the reference Brier score ~\autoref{def:briarref}; in practice, it's the improvement over this ``lazy'' approach is what matters. Thus, it's common to report the \emph{skill score} $SS$, representing the improvement over the reference Brier  value, calculated as ~\autoref{define:SS}. Skill score value ranges from $-\infty$ for model that is almost
always completely wrong in its confidence,  to $1$ for a model whose confidence values align almost perfectly
with occurrence for each event in the sample (\emph{E.g.} $100\%$ when it actually rains, and $0\%$ if not.)


\begin{equation}
acc(B_i) = \frac{1}{|B_i|} \sum_{\hat{o_i}\in B_i} {\hat{o_i}}
\label{def:accuracy}
\end{equation}

\begin{equation}
conf(B_i) = \frac{1}{|B_i|} \sum_{\hat{p_i}\in B_i} {\hat{p_i}}
\label{def:confidence}
\end{equation}
\begin{equation}
	ECE = \displaystyle\sum_{i=1}^m \frac{\mid B_i \mid}{N} \lvert \text{acc}(B_i) - \text{conf}(B_i) \rvert
    \label{def:ECE}
\end{equation}


Another way of measuring calibration is $ECE$; this measure relies on bucketing prediction probabilities of the $N$ samples in a sample $T$ into ranges (\emph{e.g.,} $\{0\text{--}0.1, 0.1\text{--}0.2, \ldots, 0.9\text{--}1\}$), and measuring the empirical frequency of the predicted outcome in each bucket. Thus, in the example, with 
well-calibrated confidences, one would expect that in the lowest bucket, the empirical frequency is about $10\%$, and the highest bucket, around $90\%$. $ECE$ is then measured
as the average difference between the average confidence  and average outcome frequency in each bucket. 
Again, we assume a sample set $T$ of predictions, split into $m$ buckets, $B_i,~i=1 \ldots m$. 
~\autoref{def:accuracy} shows the calculation of average frequency over bucket $B_i$; 
\autoref{def:confidence} is the average confidence for $B_i$; and~\autoref{def:ECE} shows the calculation of ECE over all the $m$ buckets in the entire sample of size $N = \mid T \mid$. Clearly, better calibrated
models match accuracy and confidence better, and have lower ECE values. 
However, it should be noted that ECE can give misleadingly low values
when the models suffer ``bucket collapse'', \emph{i.e.,} always produce
the base-rate confidence for every prediction; in such cases the ECE could be zero as the values all collapse in one probability bucket, but the model's confidence are of little value. 

For completeness, we also report the popular AUC-ROC. This corresponds to the probability a randomly chosen kept token/line will have higher confidence than a randomly chosen not-kept token/line. 
Thus, it is a measure of signal in the confidence technique, but not directly a measure of calibration.

In several prior works~\cite{huang2025look,spiess2025Calibration,virk2025calibration}, calibration had been
evaluated at the level of the generative problem presented to the LLM; the goal was to produce a well-calibrated confidence for the complete
(possibly long, multi-line) generated response from the model. While this is helpful, users confronted
with a long, multi-line output might wonder if some parts were more reliable than others. Thus, 
it would be desirable to have well-calibrated confidence at the line and token granularities to focus inspection efforts. Our goal here is to showcase
techniques that can produce well-calibrated confidences at completion-, line-, and token-levels. 



Next,  we  discuss a few different ways to obtain confidence measures. 

\paragraph{Confidence from Models: Inherent \& Verbalized Reflective}
Languge Models, being stochastic generators (via sampling), do inherently associate confidences with generated
output. Thus, if $n$ tokens are generated, each is associated with a confidence value directly by 
the model\footnote{This is readily obtained with open-source models; some closed-source models block access
to model confidence.}. These per-token values could be accumulated at a line, or a full problem level, to obtain confidences at varying granularities. In addition, models could also be prompted again after the initial generation, to ``reflect'' upon the generated content, and verbalize ratings that the generated output is indeed correct. Prior work~\cite{huang2025look,spiess2025Calibration} has explored some of these approaches, at the coarsest (full generation) granularity, and reported positive skill scores at that coarse-grained level. In this work, we use reflective prompting to obtain  fine-grained verbalized scores, at the line-level. 

\paragraph{Confidence from Sampling \& Differencing} 
Localized, well-calibrated confidence assessment aims to predict \emph{exactly which} parts of
generated text are low-confidence, and thus most likely to change. One way to do this is 
compare generated code with other possible versions of the same code. 
For example, if we could (hypothetically) obtain various possible future versions of the code, then, by simple differencing, we could localize the parts of code that remain unchanged across the possible futures, and which parts do not. We could then report greater confidence in the parts that remain unchanged across more of the possible futures. This is similar to the ``self-consistency'' approach of \citet{wang2022self}. Various work has deployed consistency to estimate confidence in natural language \cite{Manakul2023SelfCheckGPTZB, kuhn2023semanticuncertaintylinguisticinvariances, Farquhar2024DetectingHI}. \citet{rusure} use multiple samples from a code model to label tokens as ``unsure'' or ``sure''. Rather than a calibration focus, they frame objectives in terms of a utility function, and use a clever optimization algorithm to compare samples which is more complex than what we explore. \citet{Bhatt2025CtrlZCA} uses resampling to estimate trustworthiness of model-generated shell commands. In this work, we  generate to get
fine-grained confidences at both line- and token-levels, using multiple sample generations, as
we describe below in~\autoref{sec:multisampling}. 

\paragraph{Confidence from Latent Probing}
Generative language models have very high-dimensional, rich, internal representations of the current state of the text they are generating. Prior work~\cite{bereska2024mechanistic,bartoszcze2025representation} has shown that concepts like honesty, correctness \emph{etc} can be fairly accurately mapped out of internal representations. This relates to studies of probes for hallucination or confidence about certain non-code knowledge \cite{slobodkin-etal-2023-curious, chwang2024androidsknowtheyredreaming, ferrando2025iknowentityknowledge}. \citet{GoldowskyDill2025DetectingSD} connects probes to scheming deception in LLMs. Typically, these latent states  corresponding to generated tokens (or lines, or complete solutions)
are ``probed'', and then an extra trainable layer is trained, using  labeled data at the right
level of granularity (\emph{e.g.,} exactly which tokens, lines etc need to be repaired) to predict
granular confidences of correctness. 
Beyond fitting additional layers on the latent state, some work has explored how latent state evolves between the layers as a signal of confidence \cite{subramani2025micecatsmodelinternalconfidence}. This corpus of work suggests there is opportunity to leverage the representations for many purposes.
Some recent work has also explored this approach code. \citet{Bui2025CorrectnessAO} explored classifying the correctness of generated code with probes, but did not focus on calibration or localization. 

Most similar to our work, recently \citet{huang2025risk} explored line-level confidence estimates of code by extracting representations on newline tokens. Their probing experiments are similar to ours, but they include a representation learning pretraining stage via sparse autoencoders. They use existing datasets which make line-level repair labels
available to train a layer to produce confidences. We fully generate our own training repair labels on model-generated code (and verified with hidden test cases), using 1000+ problems which challenge even capable models like GPT-4o. 
Rather than reporting calibration measures, they rely instead on top-k recall. 
Our work also differs from~\cite{huang2025risk}
in that we include \emph{token-}level calibrated confidence, and explore different set of techniques such as multi-sampling and reflective prompting. 
We note here we find reasonable token-level calibration (Brier skill scores in excess of the 0.05 value
conventionally deemed desirable in other probabilistic forecast settings\footnote{See \url{https://www.dwd.de/EN/ourservices/seasonals_forecasts/forecast_reliability.html}}). 
We note
that the targeted sparse auto-encoding pre-training in~\cite{huang2025risk} could synergize well with our
goal of token-level prediction, which we hope to explore in future work. 


\paragraph{Confidence: Platt Rescaling}
A perfectly calibrated model exactly aligns its computed confidence with empirical likelihood of correctness. In practice, raw model confidences
may be slightly over- or under-confident, but may still have good, proportionate signal of correctness.
To improve the empirical fit, \citet{plattProbabilisticOutputsSupport1999b} introduced logistic rescaling (using just 2 parameters, for bias \& scaling) to better fit computed confidence to actual correctness. Platt scaling requires a modest amount of empirical data, but improves calibration, and we report its benefits. It should be noted that sometimes Platt scaling results
in bucket-collapse (where all predictions collapse into the same probability, giving misleadingly low ECE values). In this
case, the skill score will be poor, and thus ECE still provides a useful additional measure. 

\paragraph{Localizing}
Some works discussed above have some localizing components, but there are other related concepts. There is traditional work in software fault localization, such as spectrum-based approaches \cite{wong2016survey, de2016spectrum}. Work has extended this to prompted LLMs \cite{Kang2023AQA}. These traditionally depend on failing test cases, whereas our setting does not assume access to test cases (which are often unavailable when LLMs are used to generate \emph{di novo} code), and focuses on probabilistic estimates of confidence. \citet{Vasconcelos2023GenerationPA} conducts a human subjects study showing value in token-level confidences with labels derived from human-written patches. This result suggests that better-calibrated confidences
associated with generated code would be also valued by developers. 
\citet{palacio2024trustworthyinterpretablellmscode} looks at confidence at an AST level by aggregating the intrinsic probabilities and modeling interactions with AST node categories. 
In natural language, various works have explored extracting facts or claims from an LLM (often via another LLM query) and then estimating confidence \cite{yuan2024factlevelconfidencecalibrationselfcorrection, fadeeva2024factcheckingoutputlargelanguage}.

\section{Gathering Minimal Intent Aligning Patches}\label{sec:locdataset}
\label{sec:gathering}

We now describe our experimental method for gathering
localized correctness data for comparing localized uncertainty methods.

In our framing we have a ``model under test'' (\mutt{}) which can generate code. 
We wish to estimate localized uncertainty of lines/tokens
where its generated output probably doesn't ``match intent''. 
We use this term rather than traditional software terms such as ``buggy'' or ``defective'' for several reasons.
First, ``matching intent'' is a more general term that can apply outside of software. Second, this term captures the concept that uncertainty in the generated
code might arise from ambiguity in the prompt, rather than model capability.  For our experiments, 
we assume that ``intent" is fully defined by hidden test cases, and that intent is matched when the generated code passes all the test cases. While test sets may not in general be available
when the code is first written, they provide a reasonable basis for evaluating correctness/intent
of generated code. 

When generated code fails tests, we attempt
to find a \emph{conservative patch} that passes the failed test, and changes the generated code as little as possible. Of course, one could always change the entire generated code to
the known (correct) solution when available; but as we report later, there are often
cases where the LLM offers a near-miss solution which looks different from the known solution, and a small change to this near-miss solution is adequate. 
\dolater{For example, in our }

\subsection{Data Collection}\label{sec:collection}


\begin{table*}[htbp]
\resizebox{\textwidth}{!}{%
\begin{tabular}{l l c c c c c c c c c}
\toprule
 &  & \multicolumn{3}{c}{Solve Step} & \multicolumn{4}{c}{Patch Step} & \multicolumn{2}{c}{Final Used Data} \\
\cmidrule(lr){3-5}
\cmidrule(lr){6-9}
\cmidrule(lr){10-11}
Model Name & Dataset & Available & Pass $\checkmark$ & Fail $\times$ & \makecell{Count\\w/ Model} & \makecell{Avg Toks\\w/ Model} & \makecell{Count\\w/ Ref} & \makecell{Avg Toks\\w/ Ref} & Problems & \makecell{Total\\Tokens} \\
\midrule
gpt-4o & HumanEval+ & 164 & 139 (85\%) & 25 & 22 & 7.5 (16\%) & 2 & 25.0 (47\%) & 163 & 9k \\
gpt-4o & LiveCodeBench & 1055 & 361 (34\%) & 694 & 293 & 97.3 (46\%) & 0 & N/A & 654 & 101k \\
gpt-4o & MBPP+ & 356 & 273 (77\%) & 83 & 54 & 10.8 (18\%) & 26 & 54.6 (71\%) & 353 & 12k \\
gpt-4o & RepoCod-s & 200 & 49 (24\%) & 151 & 65 & 44.1 (28\%) & 84 & 135.4 (64\%) & 198 & 29k \\
\midrule
gpt-4o & Total & 1775 & 822 (46\%) & 953 & 434 & 74.0 (39\%) & 112 & 114.7 (65\%) & 1368 & 152k \\
\bottomrule
\end{tabular}
}%
\caption{Descriptive statistics on the dataset contents split by source dataset for GPT-4o. The 'Solve Step Pass $\checkmark$' represents a Pass@1. The 'Patch Step Count' shows the number of failing solutions successfully patched. It split by whether the patch was generated with a model, or having to fallback on a reference solution.The 'Avg Tokens' shows the number tokens not kept in the patch along with what percent of the tokens in the solution were not kept. The model generated patches are typically more more minimal. Final data shows used problems and token totals. }
\label{tab:dataset_stats_gpt_4o}
\end{table*}

\begin{table*}[htbp]
\resizebox{\textwidth}{!}{%
\begin{tabular}{l l c c c c c c c c c}
\toprule
 &  & \multicolumn{3}{c}{Solve Step} & \multicolumn{4}{c}{Patch Step} & \multicolumn{2}{c}{Final Used Data} \\
\cmidrule(lr){3-5}
\cmidrule(lr){6-9}
\cmidrule(lr){10-11}
Model Name & Dataset & Available & Pass $\checkmark$ & Fail $\times$ & \makecell{Count\\w/ Model} & \makecell{Avg Toks\\w/ Model} & \makecell{Count\\w/ Ref} & \makecell{Avg Toks\\w/ Ref} & Problems & \makecell{Total\\Tokens} \\
\midrule
Qwen2.5Coder & HumanEval+ & 164 & 112 (68\%) & 52 & 44 & 6.7 (14\%) & 7 & 29.6 (64\%) & 163 & 8k \\
Qwen2.5Coder & LiveCodeBench & 1055 & 217 (21\%) & 838 & 429 & 68.8 (43\%) & 0 & N/A & 646 & 73k \\
Qwen2.5Coder & MBPP+ & 356 & 248 (70\%) & 108 & 75 & 7.6 (24\%) & 30 & 45.1 (72\%) & 353 & 12k \\
Qwen2.5Coder & RepoCod-s & 200 & 31 (16\%) & 169 & 75 & 62.2 (38\%) & 87 & 119.0 (67\%) & 193 & 27k \\
\midrule
Qwen2.5Coder & Total & 1775 & 608 (34\%) & 1167 & 623 & 56.2 (38\%) & 124 & 96.1 (68\%) & 1355 & 120k \\
\bottomrule
\end{tabular}
}%
\caption{Descriptive statistics on the dataset contents split by source dataset for Qwen 2.5 Coder 7B Instruct. See above for more details on cols.}
\label{tab:dataset_stats_qwen7b}
\end{table*}

We first prompt the \mutt{} to generate program code for several problems. We then take the failed generated programs (with failed test cases) and prompt an ensemble of more powerful LLMs to construct a minimal fix to the program to pass the tests (``match intent"). Finally, we standardize the tokenization and calculate a diff to capture the fix.  
The diff on the minimal patch serves as the comparison ground truth for our calibrated localization measurements, by identifying which tokens are correct (remain unchanged) and which are not. 

\paragraph{Problem Sources} We use a diverse collection of function generation problems. 
This includes HumanEval+ and MBPP+ ~\cite{liuYourCodeGenerated2023a}\footnote{HumanEval+ is an enhanced version of the popular \cite{codex} dataset with more extensive test cases. Similarly, MBPP+ is a more extensively tested version of \cite{googlecodex}}, LiveCodeBench~\cite{jain2024livecodebench}, and RepoCod~\cite{repocod}.
These datasets all comprise NL-to-code generation problems. RepoCod in particular is
both realistic \& challenging: it includes a large collection of real programming tasks from popular GitHub  projects, each with an extensive collection of tests. We sample a subset of RepoCod, focusing on problems which do not require networking, are not flaky~\cite{parry2021survey} to test\footnote{though still observe some flakiness even in our sampled subset}, execute in about 3 minutes per problem, and are not excessive in size.

\paragraph{Sampling From a \mutt{}:} In this study, ``model under test'' (also referred to as ``base model'', or \mutt{}) is either GPT-4o~\cite{openai2024gpt4ocard} or Qwen2.5 Code 7B~\cite{hui2024qwen2}. For the \mutt{} we want to estimate uncertainty localized to lines or tokens. For each problem, we sample temperature zero solutions using our custom evaluation framework. Performance of GPT-4o is shown in Table \ref{tab:dataset_stats_gpt_4o} and Qwen performance is shown in Table \ref{tab:dataset_stats_qwen7b}. Note the low success rate for
the most realistic RepoCod dataset (24\% for GPT-4o and 16\% for Qwen). 

\paragraph{Finding A Minimal Patch:} After we have our generated candidate solutions (with many of them failing tests), we use a more powerful model to find patches while changing as little of the solution as possible. During this process, our prompts include ``full information'', with  details about intent that were hidden from the original solver, 
such 
the reference ``gold'' test-passing solution (if available), longer instructions, and details about failing test cases. We get repairs from each of two "Fixer" models: o4-mini~\cite{OpenAIOA} and Claude 4 Sonnet \cite{anthropic2025systemcard} under prompt variations (eg, full rewriting or more focused editing). We then filter to repairs that yield a solution which passes all the tests, and then select the solution with the smallest diff (as measured by changed or inserted lines). If the solution was originally correct, then the diff is empty. Solutions that failed and no patch is found are discarded (this is relatively rare in all datasets except
LiveCodeBench, where no reference ``gold'' solution is available to fall back to). Thus, our selected samples include: 
\begin{itemize}
\item Correct LLM-generated solutions that pass all test cases (no incorrect lines or tokens)
\item Incorrect LLM-generated solutions for which either o4-mini or Sonnet found a minimal repair (with incorrect lines or tokens that required edits in the diff)
\item Incorrect LLM-generated solutions for which the Fixing models failed, where a ``Gold'' reference solution was available  (with incorrect lines marked by diffing with the ``Gold'' solution). 
\end{itemize}

\dolater{prem Add the patches, and may be also the relative patch performance of Claude vs. GPT} 
\dolater{prem put the fix prompts here}
\dolater{david paste the prompts here in latex comment}

\paragraph{Creating Reference Error Labels:} Different \mutt{}s use different tokenizers. Since correct and incorrect tokens
must be consistently marked, we chose to normalize tokenization to Qwen2.5 tokens.
With this normalization, we calculate diffs using the python `difflib' algorithm. 
Some of the tokens ($T_s$) of each solution $s$ are then associated with a label indicating that they are edited in the minimal patch, \emph{viz.} they are incorrect. For a token to be considered ``kept'' and correct, it must both be equal to the patch token, and not have any insertions immediately before.

In our study, we measure calibrated localized uncertainty at 
several localization granularities: the token level, the line level, and the full problem level. 
In our framing, an arbitrary set of tokens (such as those within a line) can be combined where the label is defined as the logical conjunction of the individual token labels; thus a line or a solution is labeled ``correct'' exactly when all the constituent tokens are labeled ``correct''. 

\subsection{Collection Results and Discussion} Tables \ref{tab:dataset_stats_gpt_4o} and \ref{tab:dataset_stats_qwen7b} summarize some of the statistics from the data collection process for GPT-4o and Qwen2.5 Code 7B respectively. 
The two models have different initial solve rates. 
GPT-4o solves \GptNoSolvePct of problems, and Qwen2.5 solves \QwenNbSolvePct of problems, with variance depending on the complexity of the problems.
Of the failing problems, patches are found for
\ComboPatchPct of problems (considering both GPT-4o and Qwen2.5). About \ComboModelPatchPct of the patches are via patching with models, with the rest having to fall back to a reference solution if available for the dataset.
Model patches are typically much more minimal than comparing to the reference solution.

For evaluation we use 5 fold cross validation. Tables \ref{tab:main_metrics_GPT_4o} and \ref{tab:main_metrics_Qwen2.5_Coder} give average results across folds. The use of crossfold validation important when doing training-heavy techniques like probing.


\begin{table*}[htbp]
\centering
\resizebox{\textwidth}{!}{%
\begin{tabular}{l c c c c c c c c c c}
\toprule
 & \multicolumn{5}{c}{Line-Level} & \multicolumn{5}{c}{Token-Level} \\
 & \multicolumn{3}{c}{Unscaled} & \multicolumn{2}{c}{Scaled} & \multicolumn{3}{c}{Unscaled} & \multicolumn{2}{c}{Scaled} \\
\cmidrule(lr){2-4}
\cmidrule(lr){5-6}
\cmidrule(lr){7-9}
\cmidrule(lr){10-11}
Technique & BSS $\uparrow$ & ECE $\downarrow$ & AUC $\uparrow$ & BSS $\uparrow$ & ECE $\downarrow$ & BSS $\uparrow$ & ECE $\downarrow$ & AUC $\uparrow$ & BSS $\uparrow$ & ECE $\downarrow$ \\
\midrule
Token Prob & -0.34 & 0.29 & 0.61 & 0.03 & 0.01 & -0.28 & 0.24 & 0.61 & 0.02 & 0.00 \\
Multisampling & -0.13 & 0.20 & 0.72 & 0.13 & 0.03 & -0.16 & 0.20 & 0.73 & 0.13 & 0.03 \\
Line Verbalized & -0.18 & 0.26 & 0.75 & 0.17 & 0.11 & -0.11 & 0.21 & 0.72 & 0.11 & 0.09 \\
Probe-0.5B & 0.17 & 0.05 & 0.76 & 0.19 & 0.02 & 0.13 & 0.05 & 0.74 & 0.14 & 0.02 \\
Probe-7B & 0.16 & 0.10 & 0.78 & 0.22 & 0.02 & 0.14 & 0.07 & 0.76 & 0.17 & 0.03 \\
\bottomrule
\end{tabular}
}%
\caption{Calibration Results for GPT-4o with various techniques. We show both unscaled and Platt scaled results and localizing at the line and token level.}
\label{tab:main_metrics_GPT_4o}
\end{table*}

\begin{table*}[htbp]
\centering
\resizebox{\textwidth}{!}{%
\begin{tabular}{l c c c c c c c c c c}
\toprule
 & \multicolumn{5}{c}{Line-Level} & \multicolumn{5}{c}{Token-Level} \\
 & \multicolumn{3}{c}{Unscaled} & \multicolumn{2}{c}{Scaled} & \multicolumn{3}{c}{Unscaled} & \multicolumn{2}{c}{Scaled} \\
\cmidrule(lr){2-4}
\cmidrule(lr){5-6}
\cmidrule(lr){7-9}
\cmidrule(lr){10-11}
Technique & BSS $\uparrow$ & ECE $\downarrow$ & AUC $\uparrow$ & BSS $\uparrow$ & ECE $\downarrow$ & BSS $\uparrow$ & ECE $\downarrow$ & AUC $\uparrow$ & BSS $\uparrow$ & ECE $\downarrow$ \\
\midrule
Token Prob & -1.18 & 0.54 & 0.55 & 0.01 & 0.01 & -0.62 & 0.38 & 0.52 & 0.00 & 0.00 \\
Multisampling & 0.01 & 0.17 & 0.72 & 0.13 & 0.04 & 0.06 & 0.14 & 0.74 & 0.17 & 0.04 \\
Line Verbalized & -0.72 & 0.43 & 0.71 & 0.10 & 0.10 & -0.35 & 0.30 & 0.65 & 0.04 & 0.08 \\
Probe-0.5B & 0.30 & 0.05 & 0.82 & 0.31 & 0.02 & 0.20 & 0.05 & 0.77 & 0.21 & 0.02 \\
Probe-7B & 0.29 & 0.08 & 0.83 & 0.33 & 0.02 & 0.23 & 0.05 & 0.79 & 0.25 & 0.02 \\
\bottomrule
\end{tabular}
}%
\caption{Calibration Results for Qwen2.5-Coder-7B-Instruct with various techniques.}
\label{tab:main_metrics_Qwen2.5_Coder}
\end{table*}

\section{Baseline: Raw model probability}

\label{sec:baseline}

A simple measure of localized uncertainty is the token probability produced by the \mutt{}. 
To aggregate over multiple tokens in a line or a problem, we take the minimum probability of any of the tokens. We found that aggregating using the Minimum probability offered better unscaled calibration than alternative aggregations like arithmetic mean or geometric mean, though different aggregations are essentially equivalent after rescaling. When looking at a token level, it is possible the generative model might misalign with our normalized tokenization of Qwen2.5 tokens. 
We apply a greedy algorithm for approximately aligning the tokenizations. When a single normalized token covers multiple tokens from the generative model, we take the minimum probabilty of any overlapping tokens.

\paragraph{Results, Cumulative:} As in earlier works~\cite{spiess2025Calibration}, raw model probability
(see the ``Token Prob'' line in Tables~\ref{tab:main_metrics_GPT_4o}, \ref{tab:main_metrics_Qwen2.5_Coder}) is poorly calibrated at line level (negative skill score, BSS = -0.34), and
token level (BSS=-0.28); even after Platt scaling the skill score is quite low at both granularities. The low value of ECE after Platt scaling reflects
``bucket collapse'', where the rescaled predictions are all close the base-rate value. In each
partition, we train the two Platt scaling parameters over 80\% of the partition,
and apply it on the remaining 20\% of the partition; this process is done over 5 separate folds within
each partition. Thus every prediction in the dataset gets associated with 5 different scaled probabilities, and thus we get 5 different Brier scores and skill scores; we report the average. 

\begin{table*}[htbp]
\centering
\resizebox{\textwidth}{!}{%
\begin{tabular}{l l c c c c c c c c c c}
\toprule
 &  & \multicolumn{5}{c}{Line-Level} & \multicolumn{5}{c}{Token-Level} \\
 &  & \multicolumn{3}{c}{Unscaled} & \multicolumn{2}{c}{Scaled} & \multicolumn{3}{c}{Unscaled} & \multicolumn{2}{c}{Scaled} \\
\cmidrule(lr){3-5}
\cmidrule(lr){6-7}
\cmidrule(lr){8-10}
\cmidrule(lr){11-12}
\mutt & Eval Dataset & BSS $\uparrow$ & ECE $\downarrow$ & AUC $\uparrow$ & BSS $\uparrow$ & ECE $\downarrow$ & BSS $\uparrow$ & ECE $\downarrow$ & AUC $\uparrow$ & BSS $\uparrow$ & ECE $\downarrow$ \\
\midrule
GPT-4o & HumanEval+ & -0.04 & 0.04 & 0.70 & 0.01 & 0.01 & -0.35 & 0.04 & 0.66 & 0.00 & 0.00 \\
GPT-4o & MBPP+ & -0.19 & 0.19 & 0.60 & 0.02 & 0.00 & -0.17 & 0.15 & 0.56 & 0.00 & 0.00 \\
GPT-4o & LiveCodeBench & -0.28 & 0.26 & 0.59 & 0.02 & 0.01 & -0.25 & 0.23 & 0.59 & 0.01 & 0.01 \\
GPT-4o & RepoCod-s & -1.20 & 0.54 & 0.63 & 0.04 & 0.03 & -0.72 & 0.42 & 0.58 & 0.01 & 0.02 \\
Qwen2.5Coder & HumanEval+ & -0.19 & 0.16 & 0.54 & -0.01 & 0.01 & -0.08 & 0.06 & 0.51 & -0.00 & 0.00 \\
Qwen2.5Coder & MBPP+ & -0.33 & 0.26 & 0.56 & 0.01 & 0.00 & -0.19 & 0.16 & 0.51 & 0.00 & 0.00 \\
Qwen2.5Coder & LiveCodeBench & -1.40 & 0.58 & 0.54 & 0.00 & 0.01 & -0.65 & 0.40 & 0.52 & 0.00 & 0.00 \\
Qwen2.5Coder & RepoCod-s & -2.11 & 0.67 & 0.58 & 0.01 & 0.00 & -1.22 & 0.55 & 0.50 & 0.00 & 0.00 \\
\bottomrule
\end{tabular}
}%
\caption{\textbf{Token Prob} estimates split by dataset for each generating Model Under Test (\mutt).}
\label{tab:cross_se_robustness_line_token_prob}
\end{table*}
\paragraph{Results, per Dataset:}
We now describe the results broken up for each dataset. In this setting,
the raw, Brier score and ECE are calculated as before, 
separately for each dataset. Then we  calculate the skill score
using the base rate for each dataset. 
Finally, when doing Platt scaling, we use a five-fold
80-20 split \emph{within each dataset} to learn the rescaling, and calculate
the Brier score and skill score using the base rate for that dataset as before. Again, we notice that raw probabilities at both line- and token-levels aren't very well-calibrated, for any of the datasets;
and scaling provides limited help.

\section{Latent Probes of Uncertainty}\label{sec:latent_probes}
\label{sec:probing}

Language model representations enable the model to produce complex code; as discussed in Section~\ref{sec:related}, prior work has shown that latent representations can represent several pertinent phenomena.
We exploit latent representations to capture \emph{localized uncertainty} in code.

\paragraph{Probing Background:} A transformer model is a repeated stack of blocks, or layers, where at each layer, each token yields a fixed-dimensional vector encoding information about the token, previous context, and a latent view of what tokens may lie ahead. These vectors presumably represent a combination of concepts about the 
text being generated. 
A traditional approach to extracting specific concepts represented
in these vectors is ``linear probes''. 

\subsection{Enhanced Probes}

We deploy a more expressive variant of a probe model to better meet the goals of our problem. We wished to support set queries over multiple tokens (such as those within a line). We also use existing representations created by the model. 
Techniques that would modify the full parameter space and forward pass of the model are left
out of scope (such as full fine-tuning, or low-rank fine-tuning) for now.

This leaves many design choices in probing, \emph{e.g.,} where to extract embeddings, the model used for embeddings, 
complexity of a projection on that embedding, and how to aggregate the embeddings across multiple tokens. We explore several parts of this design space.
We conducted a grid search over several options, and discuss how critical the choice appears to be to calibrated uncertainty estimation. We refer to the eta-squared ($\eta^2$) \cite{cohen1988power}, which loosely estimates the variance explained by a factor, to get an approximate sense of relative importance. This is shown in Table \ref{tab:factor_importance_eta_squared} with additional visualization in the appendix.


\begin{table}[htbp]
\centering
\begin{tabular}{c c c}
\toprule
Factor & Mean BSS $\eta^2$ & Mean ECE $\eta^2$ \\
\midrule
Gen Model & \textbf{0.599} & 0.004 \\
Loss Style & 0.037 & \textbf{0.325} \\
Agg Style & 0.025 & \textbf{0.304} \\
Embedding LM & \textbf{0.186} & 0.005 \\
Aggregator & 0.058 & 0.002 \\
Hidden Dim & 0.000 & 0.001 \\
\bottomrule
\end{tabular}
\caption{A comparison of probe model design choices. Factor importance analysis showing proportion of variance explained ($\eta^2$) for Mean BSS and Mean ECE metrics. The mean BSS is averaging 6 values, the token/line/problem level BSS for both scaled and unscaled. The Mean ECE is the mean of 3 values, the unscaled ECE (we exclude scaled ECE as it almost always is near 0). Higher $\eta^2$ indicates greater factor importance.}
\label{tab:factor_importance_eta_squared}
\end{table}

\paragraph{Which model (for embedding)?} In some cases, where the generative \mutt{}
is cloud-based (or otherwise opaque) the embedding of the generative model may be unavailable. In such cases, embeddings of generated code may have to be calculated locally, on a likely lower-resource machine. Thus we might have a small auxiliary model to ``critique'' the generative model and provide confidences. For this setting we use Qwen 2.5 Coder model with 0.5B parameters to produce embeddings. We also compare to using the larger Qwen 2.5 Coder Instruct model with 7B parameters. When results from GPT-4o as the \mutt{}, we are testing the cross-model generalization. When looking at results from Qwen2.5 Code 7B as the \mutt{}, we have both a small supervisor 0.5B-parameter probe from the same model family and a probe of the embeddings from the 7B-parameter \mutt{} itself. This latter setting, where the embeddings come from the same generative model, is the more traditional probe setting.

Apart from the \mutt{} choice, we find the choice of the embedding model is the most important design choice. Larger embeddings can improve confidence signal, but can come at slight unscaled calibration reduction (possibly due to greater risk of overfitting).

\paragraph{Initial Embedding} Prior work suggests that different layers of transformers have different functions~\cite[\textit{inter alia}]{Tenney2019BERTRT, Troshin2022ProbingPM, Skean2025LayerBL}. 
While top layers focus on predicting the next token, middle layers might be more abstract, and (we hypothesize) perhaps better capture the model's local uncertainty. 
Additionally, work such as \citet{behnamghader2024llm2veclargelanguagemodels} has shown that the \textit{previous} token might be a better representation of the current token than the current token itself (as autoregressive LMs focus on predicting the future). 
Given these factors, we compare several different places to extract the embedding: \texttt{MIDDLE}, \texttt{THREE-QUARTERS}, \texttt{LAST} which embed from successively later layers of the model. We also explore \texttt{SHIFTED-THREE-QUARTERS} which learns confidence from the previous token. We also explore \texttt{COMBINED} which takes \texttt{SHIFTED-THREE-QUARTERS-LAYER} concatenated with the \texttt{MIDDLE-LAYER} of the current token. 

Embedding selection does not greatly influence BSS (Brier Skill Score)  but can have influence on ECE. We find that  \texttt{LAST} layer does indeed have the worst typical BSS and calibration. 
Using either the shifted \sfrac{3}{4} layer or the \sfrac{3}{4} layer typically works equally well. A combined representation does not appear to offer clear benefit.
\texttt{MIDDLE} achieves the best typical ECE (and possible generalization), but slightly worse typical BSS.

\paragraph{Model Design} We project the extracted embedding down to a vector of size $d_1$. 
To calculate confidence on a set of tokens, we aggregate the down-projected vectors by Max Pooling. 
We then pass this aggregated vector, through logistic regression to predict a class for the query. 
This design allows for expressive queries with minimal overhead and requiring no modifications to the generative model. We can create the down-projected representation, and then aggregate over any combination of tokens. We explore different down-projection dimensionality of either 32 or 64. Additionally, we explore instead aggregating via mean pooling, or learning a 4-head attention to take a weighted average of the embeddings\footnote{In the 4-head attention, 4 units of the down projection are used to represent attention weight, with the remaining units split evenly between the 4 heads. Note this operation is not cross-attention. We are aggregating on a set of tokens such as those in a line, and don't have query or key embeddings}. We find these model design choices to have low effect on calibration.

\paragraph{Loss Function} We weight the loss of misclassifying tokens, lines, and problems equally (ie, sum the loss function of each). We start with using binary cross-entropy to predict whether the token(s) are kept or not. 
However, prior work \cite{Mukhoti2020CalibratingDN, geng2023survey, Xia2025InfluencesOL} suggest that other loss functions might yield better calibration. Inspired by this, we also explore using Focal Loss \cite{Lin2017FocalLF}. Focus loss scales the standard $-log(p_t)$  loss,
(where $p_t$ is the computed probability of the true class---in our case, that
the token is kept) by a factor $(1-p_t)^{\gamma}$. This approach effectively re-weights the training data by emphasizing hard-to-classify examples, rather than relying on class frequency in the training data\footnote{We use the default setting of $\gamma=2$, but excluding the $\alpha$ parameter as class balance is varied between localization level}.
We explore directly optimizing for Brier Score (ie, Mean Squared Error) of probabilities. 

\paragraph{Training Procedure} We train and evaluate using 5 fold cross validation for \ProbeNumEpochs epochs using the Adam optimizer \cite{kingma2017adammethodstochasticoptimization}, dropout \cite{Srivastava2014DropoutAS} of \ProbeDropoutRate, and a default learning rate of \ProbeLearningRate. As our technique does not modify the underlying transformer model, we can precompute and cache the embeddings and fit on these cached embeddings.

Finally, we rescale the confidence using Platt rescaling, as discussed in~\autoref{sec:related}.

\paragraph{Results Cumulative:}
Results of probing are shown in Table \ref{tab:main_metrics_GPT_4o} and \ref{tab:main_metrics_Qwen2.5_Coder} with results averaged over 5 folds. In this table we show a configuration of middle layer embedding, max pooling, 32 down-projection size, and Brier score loss. This is selected not to be the exact maximum of metrics (and depending on which metric one prefers or the \mutt{}, different configurations are exactly maximal), but rather it is chosen as a sensible configuration that appears robust to other parameters. We select the lower bias option for parameters with small effect such as down-projection or embedding selection.

We find that probes with even a small supervision 0.5B parameter supervision model can achieve fairly high calibration when training/evaluating on cross-folded data of all datasets. 
For GPT-4o we observe raw skill scores of 0.17 at the line level and 0.13 at the token level, which are already quite high. With
rescaling  these improve slightly to 0.19 and 0.14 respectively. The ECE and AUC values
are also best of all the approaches. 
For Qwen2.5 Code 7B, we observe a skill score of \QwenNbProbeNbLinelevelScaledBss with an ECE of \QwenNbProbeNbLinelevelScaledEce. Thus, results are better with Qwen than GPT-4o. This might partially be because of matching the embedding to the \mutt{} better, but to large extent, is likely due to Qwen making more frequent and possibly more obvious mistakes.

We find the most critical design choice is which model \mutt{} is selected to embed for BSS. For unscaled ECE, the loss style and the aggregation method are most important.


\begin{table*}[htbp]
\centering
\resizebox{\textwidth}{!}{%
\begin{tabular}{l l l c c c c c c c c c c}
\toprule
 &  &  & \multicolumn{5}{c}{Line-Level} & \multicolumn{5}{c}{Token-Level} \\
 &  &  & \multicolumn{3}{c}{Unscaled} & \multicolumn{2}{c}{Scaled} & \multicolumn{3}{c}{Unscaled} & \multicolumn{2}{c}{Scaled} \\
\cmidrule(lr){4-6}
\cmidrule(lr){7-8}
\cmidrule(lr){9-11}
\cmidrule(lr){12-13}
\mutt & Technique & Eval Dataset & BSS $\uparrow$ & ECE $\downarrow$ & AUC $\uparrow$ & BSS $\uparrow$ & ECE $\downarrow$ & BSS $\uparrow$ & ECE $\downarrow$ & AUC $\uparrow$ & BSS $\uparrow$ & ECE $\downarrow$ \\
\midrule
GPT-4o & Probe-0.5B & HumanEval+ & -0.47 & 0.13 & 0.55 & -0.00 & 0.00 & -1.28 & 0.16 & 0.59 & 0.00 & 0.00 \\
GPT-4o & Probe-0.5B & MBPP+ & 0.06 & 0.04 & 0.67 & 0.07 & 0.01 & 0.05 & 0.03 & 0.68 & 0.06 & 0.01 \\
GPT-4o & Probe-0.5B & LiveCodeBench & 0.05 & 0.05 & 0.65 & 0.07 & 0.01 & 0.04 & 0.03 & 0.64 & 0.05 & 0.01 \\
GPT-4o & Probe-0.5B & RepoCod-s & 0.06 & 0.07 & 0.66 & 0.08 & 0.01 & 0.00 & 0.08 & 0.60 & 0.03 & 0.01 \\
GPT-4o & Probe-7B & HumanEval+ & -0.42 & 0.06 & 0.55 & -0.00 & 0.00 & -0.81 & 0.09 & 0.52 & -0.00 & 0.00 \\
GPT-4o & Probe-7B & MBPP+ & -0.15 & 0.16 & 0.57 & 0.01 & 0.01 & -0.05 & 0.08 & 0.62 & 0.02 & 0.01 \\
GPT-4o & Probe-7B & LiveCodeBench & -0.03 & 0.13 & 0.65 & 0.06 & 0.02 & 0.00 & 0.09 & 0.65 & 0.06 & 0.01 \\
GPT-4o & Probe-7B & RepoCod-s & -0.07 & 0.13 & 0.63 & 0.05 & 0.02 & -0.08 & 0.14 & 0.58 & 0.02 & 0.00 \\
Qwen2.5Coder & Probe-0.5B & HumanEval+ & -0.07 & 0.14 & 0.71 & 0.08 & 0.03 & -0.64 & 0.18 & 0.68 & 0.03 & 0.01 \\
Qwen2.5Coder & Probe-0.5B & MBPP+ & 0.02 & 0.05 & 0.67 & 0.05 & 0.02 & -0.03 & 0.08 & 0.67 & 0.04 & 0.02 \\
Qwen2.5Coder & Probe-0.5B & LiveCodeBench & -0.07 & 0.22 & 0.72 & 0.14 & 0.02 & 0.03 & 0.12 & 0.68 & 0.09 & 0.02 \\
Qwen2.5Coder & Probe-0.5B & RepoCod-s & 0.01 & 0.08 & 0.63 & 0.04 & 0.02 & -0.02 & 0.07 & 0.57 & 0.01 & 0.00 \\
Qwen2.5Coder & Probe-7B & HumanEval+ & 0.20 & 0.04 & 0.78 & 0.20 & 0.02 & -0.14 & 0.08 & 0.72 & 0.07 & 0.00 \\
Qwen2.5Coder & Probe-7B & MBPP+ & 0.01 & 0.11 & 0.72 & 0.11 & 0.03 & 0.01 & 0.05 & 0.70 & 0.07 & 0.01 \\
Qwen2.5Coder & Probe-7B & LiveCodeBench & -0.01 & 0.16 & 0.72 & 0.14 & 0.02 & 0.05 & 0.10 & 0.68 & 0.10 & 0.01 \\
Qwen2.5Coder & Probe-7B & RepoCod-s & -0.15 & 0.16 & 0.57 & 0.01 & 0.00 & -0.14 & 0.16 & 0.54 & 0.00 & 0.00 \\
\bottomrule
\end{tabular}
}%
\caption{\textbf{Probing} with a leave-one-out of each dataset for each generating Model Under Test (\mutt). So we train on the datasets not given in ``Eval Dataset''}
\label{tab:cross_se_robustness_probe}
\end{table*}

\paragraph{Results, per Dataset}
When evaluating probing for a given dataset (say RepoCod) we use the
labels in the \emph{remaining, other datasets} to train the supervision model
to output raw confidences for the held out dataset
in accordance with the labeled training data from the other datasets. These
confidences are used to calculate the raw Brier
scores; the base rate for each dataset is used to compute the raw skill score.
For Platt-scaling, we use the same five-fold 80-20 split \emph{within} each dataset, as discussed in the previous section. 
At the line-level, the skill-scores (both scaled and Platt scaled) are always
better for Qwen when using 7B model for confidences, rather than the 0.5B model. 

It's notable here that the raw
and scaled skill-score per-dataset results,  in \autoref{tab:cross_se_robustness_probe} are
so much worse than in the cumulative setting (\autoref{tab:main_metrics_GPT_4o}
and \autoref{tab:main_metrics_Qwen2.5_Coder}). Probing seems to work
well when similar datasets are in the probe training data, thus in \autoref{tab:cross_se_robustness_probe}
it performs
better for HumanEval+, MBPP+, and LiveCodeBench, since they
are similar to each other, being artificially constructed
programming problems with similar prompt structure. The worst performance,
unfortunately, is on the most practically realistic (sourced from starred GitHub projects) dataset, repocod, where 
BSS gets as high 0.08 at the line level (0.5B probe for GPT4o) and
a very weak 0.03 at the token level (same setup). 
Thus success on this style of full-file prompt/task likely needs
at least some training data in this style (future work could include more than one dataset 
in this style).
As we shall see, we
get better leave-one-out results with multisampling (\autoref{tab:cross_se_robustness_multisampling}).

\section{Localized Uncertainty via Multisampling}
\label{sec:multisampling}
Language models stochastically generate outputs, given a prompt. The possibility of \emph{variance}
between stochastic outputs, for a given prompt, can be viewed as a measure of uncertainty  (whether due to vagueness in the prompt, or noise/variation in the training data) 
that the model displays when generating the outputs. Multi-sampling approaches rest on this perspective. 

In general, these methods
work by sampling a variant set $V$  of responses $v_1, v_2, \ldots v_K$, for a given prompt $x$. Intuitively, this sample space represents uncertainty over the set of possible correct responses to prompt $x$; greater self-consistency within the $v_i$ empirically indicates higher correctness rates~\cite{wang2022self}. Later works~\cite{fadeeva2023lm, bertsch2023s, rusure} show how the \emph{degree} of consistency between a given ``sample'' output $S$ and the elements of a multi-sampled ``variant''
set $V$ provides a well-calibrated measure of confidence in $S$.


In our setting, we use a sample $S$ generated at temperature $0.0$,
and a variant set $V$, with 5 samples, generated from the model at temperature $0.8$. 
For each variant $v \in V$ we take a diff from $S$, using the Python library {\small {\tt difflib}}. For every token $t$ in $S$, we determine how often
it is in the variants $v$; if $t$ is retained in just 3 of the 5 variants $v \in V$, we calculate the
confidence in the output token $t$ as $\frac{3}{5}$. 

There are a few options for how to estimate confidence at the line-level. One can apply the same approach, and use each symbol in the diff as lines rather than tokens. However, empirically we find first doing token-level diffing, and taking an arithmetic mean of the confidence in the tokens in the line results in better calibration.

We note the sample count (5) and the higher temperature for sampling ($0.8$) are 
hyper-parameters. 


%


\paragraph{Results}

Without any access to the model internals, multisampling is able to provide 
a well-calibrated signal on localized uncertainty, but requires Platt scaling. For GPT-4o we see a line-level BSS of 
0.13
and ECE of 
0.03, and at the token-level a BSS of 0.13 and an ECE of 0.03. These skill levels
are quite good, comparable the reported values in~\cite{spiess2025Calibration} at a much coarser level.

For Qwen2.5 7B,  we see a line-level 
scaled BSS of \QwenNbMultisamplingLinelevelScaledBss and ECE of \QwenNbMultisamplingLinelevelScaledEce,
and a token-level BSS of 0.17 and ECE of 0.4.

\begin{table*}[htbp]
\centering
\resizebox{\textwidth}{!}{%
\begin{tabular}{l l c c c c c c c c c c}
\toprule
 &  & \multicolumn{5}{c}{Line-Level} & \multicolumn{5}{c}{Token-Level} \\
 &  & \multicolumn{3}{c}{Unscaled} & \multicolumn{2}{c}{Scaled} & \multicolumn{3}{c}{Unscaled} & \multicolumn{2}{c}{Scaled} \\
\cmidrule(lr){3-5}
\cmidrule(lr){6-7}
\cmidrule(lr){8-10}
\cmidrule(lr){11-12}
\mutt & Eval Dataset & BSS $\uparrow$ & ECE $\downarrow$ & AUC $\uparrow$ & BSS $\uparrow$ & ECE $\downarrow$ & BSS $\uparrow$ & ECE $\downarrow$ & AUC $\uparrow$ & BSS $\uparrow$ & ECE $\downarrow$ \\
\midrule
GPT-4o & HumanEval+ & -2.55 & 0.22 & 0.59 & 0.01 & 0.00 & -4.66 & 0.20 & 0.68 & 0.02 & 0.00 \\
GPT-4o & MBPP+ & -0.06 & 0.13 & 0.72 & 0.10 & 0.03 & -0.07 & 0.12 & 0.74 & 0.12 & 0.01 \\
GPT-4o & LiveCodeBench & -0.26 & 0.24 & 0.71 & 0.11 & 0.02 & -0.31 & 0.25 & 0.72 & 0.11 & 0.03 \\
GPT-4o & RepoCod-s & -0.02 & 0.21 & 0.75 & 0.19 & 0.06 & 0.06 & 0.15 & 0.73 & 0.17 & 0.04 \\
Qwen2.5Coder & HumanEval+ & -0.50 & 0.20 & 0.63 & 0.02 & 0.01 & -1.53 & 0.21 & 0.69 & 0.02 & 0.02 \\
Qwen2.5Coder & MBPP+ & -0.27 & 0.18 & 0.63 & 0.03 & 0.03 & -0.34 & 0.18 & 0.69 & 0.07 & 0.01 \\
Qwen2.5Coder & LiveCodeBench & -0.04 & 0.18 & 0.70 & 0.11 & 0.05 & 0.03 & 0.15 & 0.72 & 0.14 & 0.04 \\
Qwen2.5Coder & RepoCod-s & -0.08 & 0.21 & 0.71 & 0.13 & 0.06 & 0.08 & 0.14 & 0.73 & 0.16 & 0.05 \\
\bottomrule
\end{tabular}
}%
\caption{\textbf{Multisampling} split by dataset for each generating Model Under Test (\mutt).}
\label{tab:cross_se_robustness_multisampling}
\end{table*}
\paragraph{Results, per Dataset} Raw Brier scores and ECE, are calculated
separately for each dataset. Raw Skill scores are calculated using the base
rate for each separate dataset. 
The Platt scaling is done as before, on a 80-20 split for each evaluation dataset, and the average skill score and ECE over five folds is reported. 
We note the best skill scores here, of all the approaches, for the most practically relevant dataset (repocod), reaching 0.19 at the line level, and 0.17 and the token level after scaling, which are very high scores, rivaling test results (at a coarse,
full-generation
level) reported previously in related work~\cite{spiess2025Calibration}. It also
performs well for LiveCodeBench, and worse for others. This is perhaps because
the models are trained on GitHub, and multisampling is a good way to capture
what they know (and don't know~\cite{kadavath2022language}) about what code to
generate in a given context. 

\section{Reflective Uncertainty}
\label{sec:reflective}

\subsection{Reflective Method}
An alternative approach to estimating uncertainty might be to have the original \mutt{} model reflect on where its own outputs are wrong and output confidence estimates. We prompt the model to estimate the confidence every line by "verbalizing" \cite{linTeachingModelsExpress2022b} its confidence as a list of float values between 0 and 1. Additional prompt details supplementary appendix.

Our prompt allows for chain of thought reasoning before the final answer, with GPT-4o typically producing avg \hardcode{432} (stddev \hardcode{244}) tokens in its response, and Qwen2.5 typically producing avg \hardcode{385} (stddev \hardcode{204}) tokens. 

This approach only solicits reflective confidence at a line-level. To evaluate at a token-level confidence, we ``deaggregate'' assigning the line-confidence to all tokens in the line. Future work could look to better refine the token estimates with reflection.

Responding to the prompt requires outputting a list of float values of the right length and format. We find that GPT-4o has about a \hardcode{96\%} prompt compliance rate, and Qwen2.5 has about a \hardcode{98\%} prompt compliance rate. In cases where response is not compliant (and thus not parsable), we fallback to estimating the base rate of the dataset.

\subsection{Reflective Results}
\paragraph{Results, Cumulative:}
Unscaled (\autoref{tab:main_metrics_GPT_4o}, \autoref{tab:main_metrics_Qwen2.5_Coder}), the verbalized line-level confidence is not well calibrated, with GPT-4o having \GptNoLineVerbalizedLinelevelUnscaledBss BSS and \GptNoLineVerbalizedLinelevelUnscaledEce ECE. This is worse in Qwen2.5 Code 7B with \QwenNbLineVerbalizedLinelevelUnscaledBss BSS and \QwenNbLineVerbalizedLinelevelUnscaledEce ECE. Platt rescaling improves calibration BSS to low positive values, but ECE remains stubbornly high.


This suggests one should not trust the raw verbalized estimates for localized uncertainty from even competent models like GPT-4o, but it can be improved with a small sample of labeled data for rescaling. Being able to tune a probe model can still give better results than just asking the model.

\begin{table*}[htbp]
\centering
\resizebox{\textwidth}{!}{%
\begin{tabular}{l l c c c c c c c c c c}
\toprule
 &  & \multicolumn{5}{c}{Line-Level} & \multicolumn{5}{c}{Token-Level} \\
 &  & \multicolumn{3}{c}{Unscaled} & \multicolumn{2}{c}{Scaled} & \multicolumn{3}{c}{Unscaled} & \multicolumn{2}{c}{Scaled} \\
\cmidrule(lr){3-5}
\cmidrule(lr){6-7}
\cmidrule(lr){8-10}
\cmidrule(lr){11-12}
\mutt & Eval Dataset & BSS $\uparrow$ & ECE $\downarrow$ & AUC $\uparrow$ & BSS $\uparrow$ & ECE $\downarrow$ & BSS $\uparrow$ & ECE $\downarrow$ & AUC $\uparrow$ & BSS $\uparrow$ & ECE $\downarrow$ \\
\midrule
GPT-4o & HumanEval+ & 0.37 & 0.04 & 0.94 & 0.36 & 0.02 & -1.10 & 0.06 & 0.90 & 0.06 & 0.00 \\
GPT-4o & MBPP+ & 0.23 & 0.11 & 0.73 & 0.21 & 0.16 & 0.06 & 0.11 & 0.65 & 0.12 & 0.13 \\
GPT-4o & LiveCodeBench & -0.12 & 0.23 & 0.75 & 0.16 & 0.12 & -0.07 & 0.18 & 0.73 & 0.11 & 0.09 \\
GPT-4o & RepoCod-s & -0.98 & 0.49 & 0.63 & 0.03 & 0.07 & -0.52 & 0.36 & 0.60 & 0.03 & 0.02 \\
Qwen2.5Coder & HumanEval+ & 0.09 & 0.13 & 0.79 & 0.19 & 0.08 & -0.51 & 0.08 & 0.79 & 0.18 & 0.02 \\
Qwen2.5Coder & MBPP+ & 0.36 & 0.09 & 0.83 & 0.39 & 0.04 & 0.06 & 0.08 & 0.76 & 0.17 & 0.03 \\
Qwen2.5Coder & LiveCodeBench & -0.82 & 0.46 & 0.73 & 0.12 & 0.09 & -0.31 & 0.29 & 0.66 & 0.06 & 0.07 \\
Qwen2.5Coder & RepoCod-s & -1.70 & 0.61 & 0.60 & 0.00 & 0.03 & -1.00 & 0.51 & 0.55 & -0.00 & 0.00 \\
\bottomrule
\end{tabular}
}%
\caption{\textbf{Line Verbalized} split by dataset for each generating Model Under Test (\mutt).}
\label{tab:cross_se_robustness_line_verbalized}
\end{table*}
\paragraph{Results, per Dataset}
These are calculated and reported as in \autoref{sec:baseline} and \autoref{sec:multisampling}. 
The results are very good here for the artificial datasets (HumanEval+, MBPP+,
and LiveCodeBench) but quite poor for the more complex repocod dataset.

We conclude with the observation that there is also a large design space of reflective techniques (possibly including supervised or RL tuned reflection), which could also be explored further in future work for localized uncertainty.
Likely with iteration it could be competitive even for more complex datasets.

\vspace{0.25in}
\textbf{Overall observation:}
Probing can work, but might need specific data. Reflection works well more simple datasets, but additional work is needed for more complex datasets (prompt iteration or specific fine tuning). We note that multisampling offers a good (and potentially usable) performance on datasets like RepoCod. 

%
%

\section{Generalization to non-Code Text}\label{sec:generalization}

Recent works \citet{Betley2025EmergentMN, Wang2025PersonaFC} reveal ``emergent misalignment'' where tuning models on buggy code produces generally misaligned models (such as producing hate-filled speech). 
This highlights the potential interrelatedness of concepts within LLMs, which might present opportunities for easier oversight. Unlike long natural language answers, code has clearer definitions of correctness via test cases. 
Thus, while our study is focused on code, we are interested in how well these techniques generalize to other domains.
We explore how well probes trained \textit{only on code} generalize to other domains.

A common problem with LLMs is ``hallucination'' where plausible-sounding, but incorrect, content is generated. 
We study whether the code model probes can
detect natural language hallucinations.

We use HaluBench \cite{Ravi2024LynxAO}, a collection of LLM hallucinations from a mix of domains to evaluate this. We sample 1200 samples from the dataset. We then format the dataset response with an ``Answer: '' prefix. We then calculate a ``problem-level'' confidence via a Qwen2.5 Coder 0.5B parameter model probe tuned on our GPT-4o code patches dataset. For the base probabilities we apply Platt scaling fit to the code data.


\begin{figure}[ht]
    \centering
    \begin{subfigure}[t]{0.28\textwidth}
        \centering
        \includegraphics[width=\textwidth]{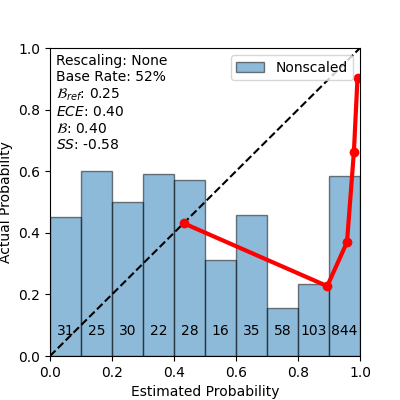}
        \caption{Unscaled}
        \label{fig:calibration_halu_unscaled}
    \end{subfigure}
    \begin{subfigure}[t]{0.28\textwidth}
        \centering
        \includegraphics[width=\textwidth]{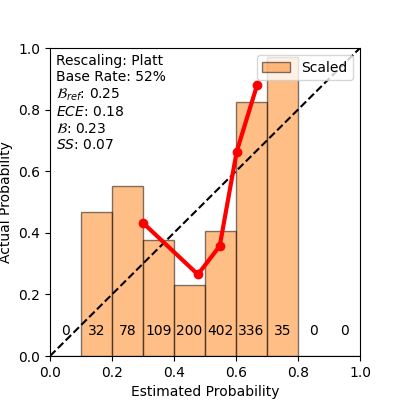}
        \caption{Scaled}
        \label{fig:calibration_halu_scaled}
    \end{subfigure}
    \caption{Calibration reliability curves of the transfer of a probe trained on code data to HaluBench. Plots show a comparison of the actual probability compared to the predicted probability. The lines plot show quintiles of the data, as predictions are not evenly distributed between buckets.}
    \label{fig:calibration_halu}
\end{figure}

A calibration curve is shown in Figure \ref{fig:calibration_halu}. Base calibration is overconfident. However, there is signal in the measure (AUC 0.72), and Platt rescaling using cross-folded HaluBench improves calibration to a skill score of 0.07.

We find that when using higher layers or the shifted/combined representation there is degraded generalization, suggesting different representations are learned in this case. We are interested in exploring this phenomenon further in future work.

While the calibration is much weaker than testing in-domain on code, we find it interesting that a probe trained only on code can detect hallucinations in natural language.

\section{Discussion}

\subsection{Design Considerations: Data, Cost, Latency}
Differing techniques showed differing calibration performance depending on context. However, it is not just calibration scores that matter.
Most human-AI collaboration currently occurs in a tight interactive loop (
\emph{e.g.}, chatbot interface, auto-complete, or in-IDE AI code-change proposals)\footnote{Although 
less tight-loop, agentic interfaces are rapidly becoming more popular.}. 
In interactive situations, \emph{latency} matters, and computational overheads
and data requirements are important considerations.
Some methods are highly parallelizable; \emph{e.g.}  multi-sampling from a model can be done in parallel if sufficient compute is available. Other methods are inherently serial---reflective prompting must be done only after initial generation. While this can increase latency, clever HCI design (such as showing first the output, and then the confidence slightly delayed) can perhaps render the increased latency less disruptive.

The \emph{cost} of a given approach could depend on specific deployments. Some LLM services charge more for multi-sampling, while others support shared cost of the input tokens. When using internal probes from the generative model itself, there is low increase in latency or cost. If using a small,
locally-running supervision model to calculate confidences (using probing), in conjunction with inaccessible cloud model, latency and cost are typically still dominated by the initial generation, but will require local resources.

\paragraph{Data Labeling} The probing approach relies on training a logistic
regression from high-dimensional internal embeddings; thus probing carries the highest  data burden of  requiring labels (correct/incorrect) at the token/line/problem level. At the per-token granularity, one typically does
have a lot of data.
Data requirements could be mitigated \emph{via} noisier representation learning (eg, the use of sparse
auto-encoders~\cite{huang2025risk} or dense representation learning), suitably adapted for our per-token setting. 
The other
approaches only require a few labels to optimize the two (bias + scaling) parameters
to fit the Platt rescaling. 

\subsection{Limitations}

\paragraph{HCI Considerations} We empirically show that several methods can provide
localized, well-calibrated confidence signals at high-levels of skill. However, these
confidence signals must be interactively displayed to developers along with generated
code; while prior work in this area exists~\cite{Vasconcelos2023GenerationPA, devic2025calibrationcollaborationllmuncertainty}  further study is needed. 

\paragraph{Performance on the Edge of the Frontier} The base models we study (GPT-4o and Qwen2.5 Code 7B) are capable, but not quite frontier models at the time of the study in early/mid 2025. 
We are constrained by practicalities around cost and scope.
Independent of cost, in this setting we must strike a balance between models being complex enough to be interesting and relevant, 
and room that we are able to repair the failing cases with even more powerful models.
In our findings, techniques like probing demonstrate promising results with an order $10^3$ data examples. 
With the budget of a large group, this is a quantity where expert human labeling is possible, thus possibly allowing for labeling of problems only expert humans can solve\footnote{Thus closer to the label of 10s of thousands of examples in ``process supervision'' \cite{Lightman2023LetsVS}}. 

A key challenge of scalable oversight \cite{Amodei2016ConcretePI, Bowman2022MeasuringPO} is eventually we will be supervising problems that are beyond human capabilities. 
In many case (such as in-domain), we find very small models (0.5B) can be competent probes of code generated with models over 10x-100x the size. This is encouraging for scalable oversight, as it implies we might be able to use much smaller models for oversight despite it not being able to solve almost any of the problems itself. 
While we should be cognizant that the performance of techniques might change in different model regimes, we believe our results are a novel and useful study of this current regime, and points to many areas for further research.

\paragraph{Overfitting Risks} Parts of our experiment protocol, such as the lack of a clean heldout test set, creates potential risk that results reflect overfitting, in particular for high bias techniques like probing with many hyper parameters. As discussed in Section \ref{sec:latent_probes}, findings do not appear to be tightly dependent on exact hyper parameters. As discussed in Section \ref{sec:generalization}, in some cases results can generalize to completely new domains if new scaling is allowed. However, as shown in the leave-one-out per dataset discussion, full robustness remains a challenge. Future work could further explore generalization and robustness.

\paragraph{Multiple Settings With Varied Results} 
Different styles (like the function generation vs repo-level generation) makes interpretation
of results less consistent.

\paragraph{Minimal Patches are Approximate} Our process of finding minimal patches is not perfect. Indeed, for general programs we might speculate it is uncomputable to know if a given patch is minimal. Thus, results are only approximate, and can be further improved with successively better techniques for patching.

\section{Conclusion}

The emergence of the use of LLMs creates many challenges for use. When one is able to construct a dataset of patches, supervised probes can give a localized signal of confidence. Our method of
enhanced probes allows joint learning of at the token, line, and problem levels.
When training is not possible, multisampling or reflective prompting is viable, but at least some data is needed for rescaling in order to achieve good expected calibration error.
Training just on source code can be a step in getting signal on even non-code edits.

The source code and dataset of our minimal intent aligning patches are available on GitHub\footnote{\url{https://github.com/DaiseyCode/LocalCalibrationPubExps}}.
We hope this work can aid in understanding of safe and reliable use of LLMs.

\bibliographystyle{ACM-Reference-Format}
\bibliography{aaai2026}

\clearpage
\appendix



\begin{table}[htbp]
\begin{tabular}{c c c c c c}
\toprule
 & \multicolumn{3}{c}{Unscaled} & \multicolumn{2}{c}{Scaled} \\
\cmidrule(lr){2-4}
\cmidrule(lr){5-6}
line\_agg & BSS $\uparrow$ & ECE $\downarrow$ & AUC $\uparrow$ & BSS $\uparrow$ & ECE $\downarrow$ \\
\midrule
mean & 0.03 & 0.14 & 0.73 & 0.14 & 0.04 \\
gmean & -0.14 & 0.20 & 0.72 & 0.13 & 0.03 \\
min & -0.28 & 0.27 & 0.72 & 0.14 & 0.02 \\
pre\_min & -0.38 & 0.30 & 0.72 & 0.13 & 0.02 \\
pre\_mean & 0.03 & 0.14 & 0.73 & 0.14 & 0.04 \\
\bottomrule
\end{tabular}
\caption{Multisampling line-level results for different line aggregators}
\label{tab:dataset_stats}
\end{table}

\begin{figure*}[htbp]
    \centering
    \includegraphics[width=\textwidth,height=0.9\textheight,keepaspectratio]{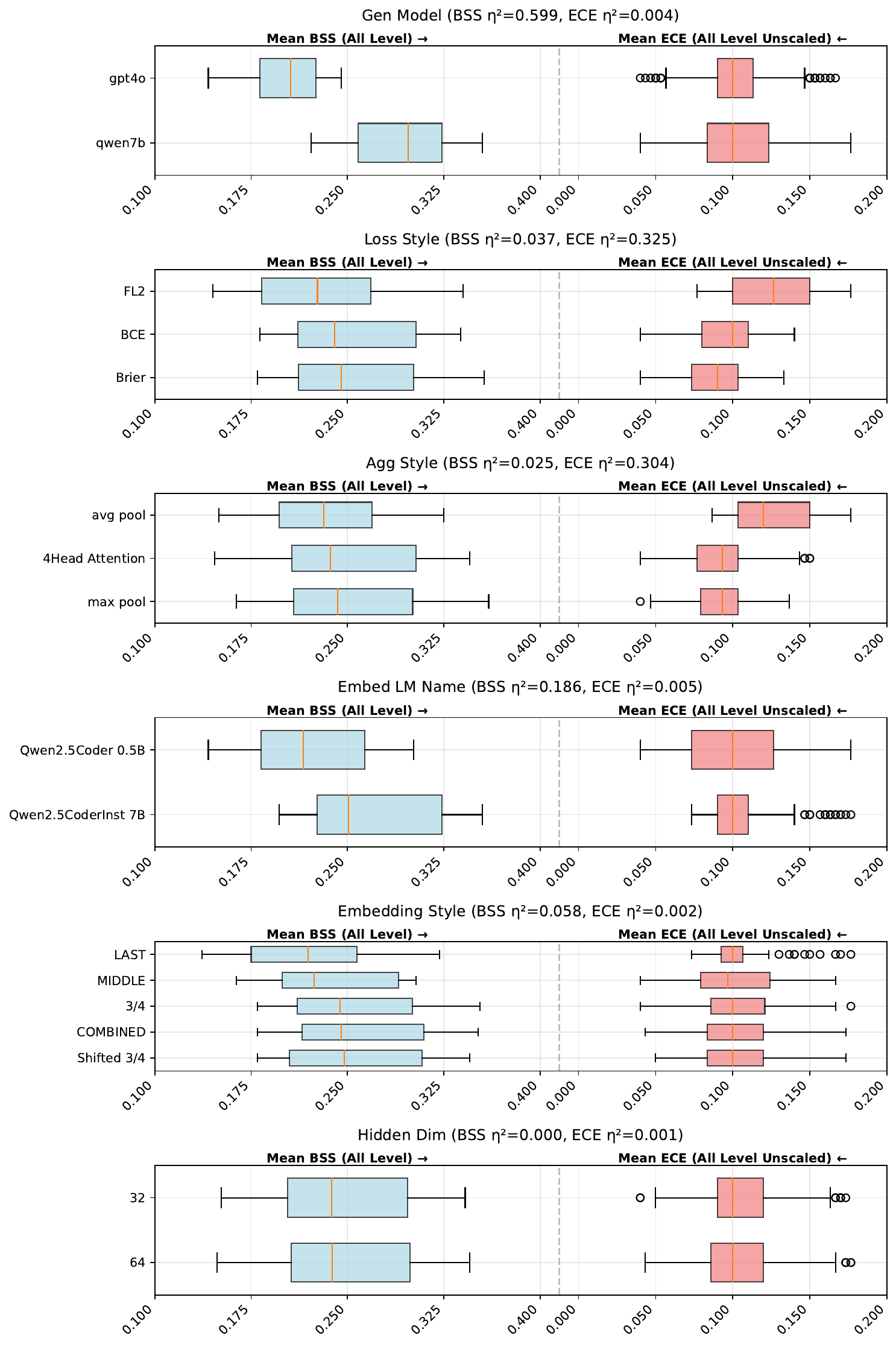}
    \caption{Factor breakdown exploring some of the design space when probing for local uncertainty.
    We gridsearch over options. Box plots show metrics accross different configurations, and help understand both the best achieved results (top of the range), but also how robust the selection of the parameter is to other choices (via the median and spread).}
    \label{fig:factor_breakdown}
\end{figure*}

\section{Reflective Uncertainty Prompt}
The following prompt template 
is used to prompt the model to verbalize its confidence.
We want to have a reasonably sized prompt, while being
specific about the lines it needs to consider confidences
for.

\begin{lstlisting}[captionpos=b, caption=Reflective uncertainty prompt construction, label=lst:reflective_prompt, basicstyle=\scriptsize\ttfamily, frame=single]
We are attempting estimated calibrated probabilities that lines of code are correct or if they will need to be edited.
Consider the following code:
```python
{content}
```
We are attempting to estimate the probability that each line of code is correct.
Please provide your estimate as a list[float] where each element is between 
0 and 1 representing the probability that the line will be correct and 
unedited. One or two digits of precision is fine.
This is the individual line probabilities so the sum is not expected to be 1.
These are the line splitting we are using:
[
    {list_of_considered_lines}
]
Create a calibrated estimate of the probability that each line is correct. 
You can consider any potential issues with the code, but then place your final answer in a markdown code block with only a list[float] of length {len(lines)}. Do not end early, and do not stop until you list all {len(lines)} probabilities corresponding to the given splits.
\end{lstlisting} 

\end{document}